\documentstyle[preprint,aps]{revtex}
\tightenlines
\newcommand{\be}{\begin{equation}}
\newcommand{\ee}{\end{equation}}
\newcommand{\s}{\cite}
\newcommand{\la}{\label}
\newcommand{\ber}{\begin{eqnarray}}
\newcommand{\eer}{\end{eqnarray}}
\newcommand{\nn}{\nonumber}
\begin{document}
\draft

\title{Method of integral transforms for few--body reactions}

\author{V.D. Efros}
\address
{
Russian Research Centre "Kurchatov Institute",
Kurchatov Square 1,\\ 123182 Moscow, Russia\thanks{email:
efros@mbslab.kiae.ru}}

\date{Received December\qquad , 1998}

\maketitle

\begin{abstract}
The method of integral transforms is reviewed. In the framework
of this method reaction observables are obtained with the bound--state
calculation techniques.  New developments are reported.
\end{abstract}
\bigskip

\vfill\eject
\section{Introduction}
The approach discussed in the paper simplifies solving reaction
problems. In particular, substantial simplifications occur at
intermediate energy when many reaction channels are effective.
No
specification of reaction channels at solving the dynamic equations is
required, and to obtain reaction observables
only the bound--state type calculations are to be performed. One
avoids the consideration of the complicated configuration--space
asymptotic behavior of continuum wave functions or the complicated
momentum--space singularities.
In case of
perturbation--induced inclusive reactions the cross section
is obtained in a closed form, while
the summation of contributions from various
final states with the same energy is avoided.
The long--range Coulomb final state interaction is
automatically taken into account in this case.

In \cite{efr80} it was suggested to compare with experiment the
Stiltjes
transform of response functions for perturbation--induced reactions,
such as the $(e,e')$ reaction. The transform is directly obtained
with the bound state few--body methods. A qualitative study of the
Fourier
transform of $(e,e')$ response functions was performed at the same
time \cite{ros,KS81} but with no intention for a complete
quantitative
evaluation. In \s{efr85} the general procedure
presented below was formulated for perturbation--induced inclusive
and exclusive reactions, and for nonperturbative processes.
Despite the successful solution of simple models, use of real
values of the transform variable in that formulation imposed too high
requirements to precision of the transform
\s{ELO93}. In \s{ELO94} the real transform variable was
changed to
the complex one which substantially improved the accuracy of the
inversion.
The application of the method separately to
various
multipoles also increased the accuracy \s{ELO94}.  Successful
applications to
three-- and four--nucleon $(e,e')$ and
total photodisintegration cross sections were given in
\s{MKOL95,ELO97A,ELO97B,ELO97C,ELO99,ELOT99}, and
the four--nucleon spectral function was obtained along
these lines
\s{ELO98}. In the response function case, a similar procedure accompanied 
with the statistical averaging is
attempted in condensed--matter physics, see e.g. \s{gub91}. 
Different from the transforms considered below it is
based on the Laplace transform calculable with the Monte--Carlo path
integration which reliable inversion is a very difficult
task.

\section{Outline of the method}

Let us consider "$E$--overlaps" defined as
\begin{equation}
r(E)=
\int d\gamma\langle\chi_2|\Psi_\gamma\rangle\langle
\Psi_\gamma|\chi_1\rangle
\delta(E_\gamma-E)\, \label{gr}
\end{equation}
where the state $\chi_1$
(but not necessarily $\chi_2$) is
localized, i.e. its
conventional norm is finite.
The functions $\Psi_\gamma$ are solutions  to
the Schr\"odinger equation:
$(\hat{H}-E_\gamma)\Psi_\gamma=0$, where $\hat{H}$ is the
Hamiltonian of a system.
The set $\Psi_\gamma$ is supposed to be complete and orthogonal.
The notation
$\int d\gamma$ implies an integration over the continuum states
plus summation over
the bound states in the set. We thus have
$\int d\gamma |\Psi_\gamma\rangle\langle\Psi_\gamma|=1$.
The $E$--overlaps (\ref{gr}) are the basic quantities of the
approach, and in this sense they replace continuum wave
functions the conventional approach deals with.

The main
points are the following.  1. Reaction observables
are expressed in terms of the $E$--overlaps $r(E)$. 2. The latter
are
computed indirectly\footnote{The
functions
$\Psi_\gamma$ are not used for obtaining $r(E)$, and
continuum wave functions thus
do not enter the calculation at all.}
with the bound state calculation techniques.
In the next section we explain the first of these points, while
in Sec.
4 and 5 the second one is explained.
Few--body calculations  done so far with the method are
outlined in Sec. 7. In Sec. 6 other non--conventional
approaches having some common features with the present
one are discussed.

Eq. (\ref{gr}) includes the case of response functions for
perturbation--induced reactions.
They arise if one
sets in (1)
$\chi_1=\hat{O}_1\Psi_0$, $\chi_2=\hat{O}_2\Psi_0$,
and $E=E_0+\epsilon$,
where $\hat{O}_{1,2}$ are
the transition
operators, and
$\Psi_0$ and $E_0$ are the
initial--state wave function and energy:
\ber
R(\epsilon)=(1/2)(x+x^*)\,,\nonumber\\
x=\int d\gamma\langle\hat{O}_2\Psi_0|\Psi_\gamma\rangle
\langle\Psi_\gamma|\hat{O}_1\Psi_0\rangle
\delta(E_\gamma-E_0-\epsilon)\,. \label{r}
\eer
In Eq. (\ref{r}) $\Psi_\gamma$ and $E_\gamma$ are the final state
wave functions and
energies, respectively. The  $\hat{O}_1=\hat{O}_2$ case is typical.

The method proceeds in the following steps \cite{efr85}.
First
some integral transform with a smooth kernel $K$
of the $E$--overlap $r(E)$, (or of the response $R(\epsilon)$),
\begin{equation}
\int K(\sigma,E)r(E)dE=\Phi(\sigma)\,, \label{tra}
\end{equation}
is calculated instead of $r(E)$ itself. To this aim,
let us multiply Eq. (\ref{gr}) by
$K(\sigma,E)$ and integrate over $E$. One obtains
\begin{eqnarray*}
\Phi(\sigma)=\int d\gamma
\langle\chi_2|\Psi_\gamma\rangle K(\sigma,E_
\gamma)
\langle\Psi_\gamma|\chi_1\rangle  \\
=\int d\gamma \langle\chi_2|K(\sigma,\hat{H})\Psi_\gamma\rangle
\langle\Psi_\gamma|\chi_1\rangle\,.
\end{eqnarray*}
Using the closure
property one gets
\begin{equation}
\Phi(\sigma)=\langle\chi_2|K(\sigma,\hat{H})\chi_1\rangle\,.
\label{phi}
\end{equation}
Eq. (\ref{phi}) may be viewed as a generalized sum rule depending on a
continuous parameter.
The expression (\ref{phi}) is evaluated with {\em the bound--state
calculation techniques} as described in Sec. 4. At the second step,
Eq. (\ref{tra})
is considered as an integral equation to invert and thus one gets
$r(E)$
from $\Phi(\sigma)$. In case (\ref{r}) this
provides
a final result since the response function is an observable
quantity.
In the general case reaction observables are
expressed as some quadratures
in terms of $E$--overlaps (\ref{gr}) thus obtained,
see the next section. (The reader interested only in the
calculation of response functions may skip the
next section except, perhaps, Eqs. (\ref{spf})--(\ref{wjac}).
Note also that in the response function case the transform variable is
changed
to $\epsilon=E-E_0$.) In the non--relativistic case,
all the considered states, or operators
may be viewed
as "intrinsic" ones.

\section{Reaction observables in terms of $E$--overlaps}

First let us consider the case of transitions to continuum
induced by
a perturbation $\hat{O}$. The matrix element
\be
M_{fi}=\langle\Psi^-_f(E)|\hat{O}\Psi_0\rangle\,, \la{mfi}
\ee
is to be calculated where $\Psi_0$ is a localized initial state
wave function, while $\Psi^-_f$
belongs to continuum. We set \s{GW64}
\be
\Psi_f^-(E)=\phi_f(E)+(E-\hat{H}-i\eta)^{-1}\bar{\phi}_f(E)\,.
\la{psim}\ee
Here $\phi_f$ is either the "channel" function that consists of
a product of
wave functions of fragments
pertaining to a given reaction
channel\footnote{The quantity (\ref{pb}) equals $V_f\phi_f$
in this case.}, or
the sum over such channel functions corresponding to (anti)symmetrization
\s{GW64}. In the normal non--relativistic case one may
use intrinsic channel functions that are angular momentum coupled
products of intrinsic wave functions of fragments and of the relative
motion function with a given orbital momentum. The functions $\bar{\phi}_f$
are
\be
\bar{\phi}_f=(\hat{H}-E)\phi_f\,. \la{pb}
\ee
Using Eq. (\ref{psim}), we rewrite (\ref{mfi}) as
\be
M_{fi}=\langle\phi_f(E)|\hat{O}\Psi_i\rangle+
\langle\bar{\phi}_f(E)|(E-\hat{H}+i\eta)^{-1}\hat{O}\Psi_i
\rangle\,.\la{m1}
\ee
The first, Born, term in (\ref{m1}) is
computed directly. The second term takes into account the final
state interaction.
We present this term in the form
\ber
\sum_n (E-E_n)^{-1}F_{fi}(E_n)+\int_{E_{th}}^\infty
(E-E'+i\eta)^{-1}F_{fi}(E')dE'\nn\\
=\sum_n (E-E_n)^{-1}F_{fi}(E_n)-i\pi F_{fi}(E)+
P\int_{E_{th}}^\infty(E-E')^{-1}F_{fi}(E')dE'\,,\la{st}
\eer
where the form factor $F_{fi}$ is defined as
\be
F_{fi}(E')=\int d\gamma\langle\bar{\phi}_f(E)|\Psi_\gamma\rangle
\langle\Psi_\gamma|\hat{O}\Psi_i\rangle\delta(E_\gamma-E')\,.
\la{ff}
\ee
Eq. (\ref{ff}) is a special case of Eq. (\ref{gr}), and the reaction
amplitude (\ref{mfi})
is thus expressed in terms of a simple quadrature (\ref{st}) over
a $E$--overlap \s{efr85}.

The above formulation is applicable when the long--range
Coulomb interaction of fragments in a final state is neglected.
In order to take into account this interaction one can proceed as follows.
To
simplify
notation we omit (anti)symmetrization of the channel wave functions.
(Changes to include (anti)symmetrization are obvious.)
Let us write
$\hat{H}=T_f+V_f$ where $V_f$ is the interaction of fragments in the
final state. Let us subtract from $V_f$ a part $U^C_f$ that includes the
long--range Coulomb
interaction: $V_f=\bar{V}_f+U^C_f$. Let us denote
$\varphi_f^-$ the "plain wave plus ingoing wave" continuum wave
functions of the Coulomb--like Hamiltonian
$\hat{H}_f^C=T_f+U^C_f$. Instead of Eq. (\ref{psim}) let us use
(cf. \s{GW64}, Chap. 5, Eq. (108))
\[
\Psi_f^-=\varphi_f^-+(E-H+i\eta)^{-1}\bar{V}_{f}\varphi_f^-.
\]
One can proceed as above with the replacement of the channel function
$\phi_f$ by the "Coulomb" channel function $\varphi_f^-$, and
 $\bar{\phi}_f$ by $\bar{V}_f\varphi_f^-$, or, more generally, by
 $(\hat{H}-E)\varphi_f^-$ if $\varphi_f^-$ is (anti)symmetrized. This
 procedure requires constructing $\varphi_f^-$ which can be done
 exactly when the Coulomb interaction in only one pair of fragments
 is present.

The arguments similar to those presented above are
valid if one replaces $\hat{O}\Psi_0$ with any other localized state.
Consider, for example, a breakup reaction amplitude in the
framework of the
Glauber approximation. It is given by the
matrix element
of the (approximate) fixed--center scattering amplitude
between the initial and
final state of the target.
We thus
encounter
again a matrix element of the form $\langle\Psi_f|\chi\rangle$,
$\Psi_f$ and $\chi$ being the final state wave function and a known
localized function, respectively. Proceeding as above in connection
with
Eq. (\ref{mfi}), and putting $\chi$ instead of $\hat{O}\Psi_0$ one
replaces the
calculation of the continuum wave function $\Psi_f$ by that of
a $E$--overlap.

In Ref. \s{ELO98} the spectral function of an $A$--body system
is expressed in
terms of an $(A-1)$--body $E$--overlap. Below the formulas are
presented in more detail. 
The spectral
function
$P(k,E)$ represents the joint probability of finding a nucleon
with momentum ${\bf k}$ and a residual subsystem of $A-1$ nucleons
with energy $E$ in the ground state $\Psi_0^A$ at rest. The energy $E$ is
calculated with respect to the ground state energy:
\be
P(k,E)=(2J_0+1)^{-1}\sum_{M_0=-J_0}^{J_0}\sum_{s_z=\pm1/2}\int dfd{\bf K_0}
\left|\langle\psi_{f{\bf K_0}}^{A-1};
{\bf k},s_zt_z|\Psi_0^A(J_0M_0)\rangle\right|^2
\delta(E-(E_f^{A-1}-E_0^A)).
\la{spf}\ee
Here $s_z$ and $t_z$ are the third components of the nucleon spin
and isospin, $E_f^{A-1}=M_f^{A-1}-(A-1)M$, $E_0^A=M_0^A-AM$, where
$M_f^{A-1}$ and $M_0^A$ are the masses of the final residual state
and the initial state, and $M$ is the nucleon mass. The functions
$\psi_{f{\bf K_0}}^{A-1}$ form the complete set of eigenstates of
the $A-1$ subsystem with the total momentum ${\bf K}_0$:
\be
\psi_{f{\bf K_0}}^{A-1}=\int d{\bf k}_1\ldots d{\bf k}_{A-1}
\varphi_{f}^{A-1}({\bf k}_1,\ldots,{\bf k}_{A-1})
\delta(\sum_{i=1}^{A-1}{\bf k}_i -{\bf K}_0)
|{\bf k}_1\ldots {\bf k}_{A-1}\rangle. \la{a1} \ee
The ground state can be written as
\be
\Psi_0^A=\int f({\bf P})d{\bf P}\int
d{\bf k}_1\ldots d{\bf k}_{A}
\varphi_0^{A}({\bf k}_1,\ldots,{\bf k}_{A})
\delta(\sum_{i=1}^{A}{\bf k}_i -{\bf P})
|{\bf k}_1\ldots {\bf k}_{A}\rangle, \la{a} \ee
where $f({\bf P})$ represents a narrow peak concentrated
in the vicinity of ${\bf P}=0$, and
$\int \left|f({\bf P})\right|^2d{\bf P}=1$.
In Eqs. (\ref{a1}), (\ref{a}) and below
we omit spin and isospin variables.

In terms of the wave functions entering (\ref{a1}), (\ref{a}) the spectral
function can be written as
\ber
P(k,E)=(2J_0+1)^{-1}\sum_{M_0=-J_0}^{J_0}\int df\delta(E-(E_f^{A-1}-E_0^A))
W_f({\bf k}), \la{spf1} \\
W_f({\bf k})=\left|\int d{\bf k}_1\ldots d{\bf k}_{A-1}
\varphi_{f}^{A-1\,*}({\bf k}_1,\ldots,{\bf k}_{A-1})
\varphi_0^{A}({\bf k}_1,\ldots,{\bf k}_{A-1},{\bf k_A})
\delta(\sum_{i=1}^{A-1}{\bf k}_i+{\bf k} )\right|^2 \la{w} \eer
with ${\bf k}_A={\bf k}$.

For the $A-1$ subsystem and  for the $A$ system in Eq. (\ref{w})
let us perform orthogonal
transformations  to Jacobi
momenta \[ \mbox{$\boldmath \pi$}_1,\ldots,
\mbox{$\boldmath \pi$}_{A-2}, \mbox{$\boldmath \Pi$}_{A-1}=
(A-1)^{-1/2}\sum_{i=1}^{A-1}{\bf k}_i;\qquad\mbox{and }
\qquad\mbox{$\boldmath \pi$}_1,\ldots,\mbox{$\boldmath \pi$}_{A-2},
\mbox{$\boldmath \pi$}_{A-1}, \mbox{$\boldmath \Pi$}_{A}
=A^{-1/2}\sum_{i=1}^{A}{\bf k}_i\,, \] respectively.
Here $\mbox{$\boldmath \pi$}_{A-1}
={[(A-1)/A]^{1/2}[{\bf k}_A-(A-1)^{-1} \sum_{i=1}^{A-1}{\bf k}_i]}$.
Eq. (\ref{w}) takes the form
\be
W_f({\bf k})=[A/(A-1)]^{3/2}\left|\int d\mbox{$\boldmath \pi$}_1\ldots
d\mbox{$\boldmath \pi$}_{A-2}
\tilde{\varphi}_f^{A-1\,*}(\mbox{$\boldmath \pi$}_1,\ldots,
\mbox{$\boldmath \pi$}_{A-2};-{\bf k})
\tilde{\varphi}_0^{A}(\mbox{$\boldmath \pi$}_1,\ldots,
\mbox{$\boldmath \pi$}_{A-2},\mbox{$\boldmath \pi$}_{A-1};0)\right|^2
\la{wjac}\ee
with $\mbox{$\boldmath \pi$}_{A-1}=[A/(A-1)]^{1/2}{\bf k}$. Here
$\tilde{\varphi}_f^{A-1}=(A-1)^{-3/4}\varphi_f^{A-1}$,
$\tilde{\varphi}_0^{A}=A^{-3/4}\varphi_0^A$, and
\[ \int d\mbox{$\boldmath \pi$}_1\ldots d\mbox{$\boldmath \pi$}_{A-2}
\tilde{\varphi}_{f'}^{A-1\,*}
\tilde{\varphi}_f^{A-1}=\delta(f-f'),\,\,\,\,\,
\int d\mbox{$\boldmath \pi$}_1\ldots d\mbox{$\boldmath \pi$}_{A-1}|
\tilde{\varphi}_0^A|^2=1. \]
The last
arguments of the functions $\tilde{\varphi}_f^{A-1}$ and
$\tilde{\varphi}_0^{A}$ are total momenta of the $A-1$ subsystem
and $A$ system, respectively.
In the non--relativistic limit $\tilde{\varphi}_f^{A-1}$ and
$\tilde{\varphi}_0^{A}$ cease to depend on these arguments.
From Eqs. (\ref{spf1}), (\ref{wjac}) one sees that the spectral function
is a parameter $k$ depending $E$--overlap  in the subspace of
$A-1$ particle.

The amplitudes of nonperturbative
reactions have been expressed in terms of $E$--overlaps in
Ref. \s{efr85}. Usual processes
with two fragments in an initial state and any number of
fragments in final states were considered.
Similar to Eq. (\ref{pb}) we shall use the functions
\be \bar{\phi}_i=(\hat{H}-E)\phi_i\,\,\,\,\,\,\,\,\,
\bar{\phi}_f=(\hat{H}-E)\phi_f\,,  \la{pb1}
\ee
where $\phi_i$ and $\phi_f$ are
defined in the same way as $\phi_f$ in (\ref{pb}).
 We rewrite the expression \s{GW64} for
the $i\rightarrow f$ transition $T$ matrix in terms of
$\bar{\phi}_{i,f}$:
\be
T_{fi}(E)=T_{fi}^{Born}+
\langle\bar\phi_f(E)|(E-\hat{H}+i\eta)^{-1}\bar\phi_i(E)\rangle\,,
\la{tm}
\ee
where
\[ T_{fi}^{Born}=\langle\bar\phi_f(E)|\phi_i(E)\rangle=
\langle\phi_f(E)|\bar\phi_i(E)\rangle \]
is the Born term.

We need to calculate the second term in Eq. (\ref{tm}) representing
the final--state interaction correction. Similar to Eq. (\ref{ff})
we define a form factor
\be
F_{fi}(E,E')=\int d\gamma\langle\bar{\phi}_f(E)|\Psi_\gamma\rangle
\langle\Psi_\gamma|\bar{\phi}_i(E)\rangle\delta(E_\gamma-E')\,,
\la{ff1}
\ee
and, finally, as in Eq. (\ref{st}), we express the contribution
in Eq. (\ref{tm}) we consider in terms
of the quantity (\ref{ff1}) . Again, Eq. (\ref{ff1})
is a special case of Eq. (\ref{gr}). Because there are 
two colliding fragments in the entrance
channel the function $\bar{\phi}_i$ is localized.
The method is directly applicable
when the long--range Coulomb interaction is neglected.

\section{Evaluation of the transforms and the choice of the
transform kernel}

Having expressed reaction observables in terms of $E$--overlaps,
now we consider the calculation of the latter quantities using
Eqs. (\ref{tra}), (\ref{phi}) of Sec. 2.
One possibility is the following. Consider some complete set of
localized
states and denote $P_N$ the projector onto the subspace of
the first $N$
states in the set.
As $N$ tends to infinity at some conditions one has
\begin{equation}
\langle\chi_2|P_NK(\sigma,\hat{H})P_N|\chi_1\rangle
\rightarrow \langle\chi_2|K(\sigma,\hat{H})|\chi_1\rangle
\label{appr}
\end{equation}
uniformly with respect to $\sigma$.
Therefore Eq. (\ref{phi}) can be approximated with the left--hand
side of Eq. (\ref{appr}).
Correspondingly, if one denotes by $\psi_n^{(N)}$   the states
that diagonalize
the Hamiltonian in the subspace determined by the projector $P_N$ 
and by $E_n^{(N)}$ the corresponding eigenenergies, then one has
\begin{equation}
\Phi_N(\sigma)=
\sum_{n=1}^N\langle\chi_2|\psi_n^{(N)}\rangle
\langle\psi_n^{(N)}|\chi_1\rangle K(\sigma,E_n^{(N)})\,.
 \label{phin}
\end{equation}
This expression can be used as an approximate
right--hand side in the integral equation (\ref{tra}).
Thus, the input to the integral equation is obtained with the
bound--state calculation technique.

Eq. (\ref{appr}) is valid when both $\chi_1$ and $\chi_2$
are localized, and the kernel $K$ is bounded uniformly with respect
to $\sigma$: $|K(\sigma,E)|<C$. If the latter is valid then the state
$K(\sigma,\hat{H})\chi_1$ is localized along with $\chi_1$. Indeed,
the norms of these two states are
$\int d\gamma |K(\sigma,E_\gamma)|^2|
\langle\Psi_\gamma|\phi_1\rangle|^2$ and
$\int d\gamma|\langle\Psi_\gamma|\phi_1\rangle|^2$, respectively.
Since the latter is finite, the former is finite as well.
Taking this into account, one sees that the quantities
$\langle\phi_2|Q_NK\phi_1\rangle$, $\langle\phi_2|KQ_N\phi_1\rangle$,
and $\langle\phi_2|Q_NKQ_N\phi_1\rangle$ tend to zero at
$N\rightarrow\infty$, $Q_N$ being $1-P_N$. This gives Eq. (\ref{phin}).

The exact solution with the right--hand side (\ref{phin}) is
\begin{equation}
r_N(E)=\sum_{n=1}^N\langle\chi_2|\psi_n^{(N)}\rangle
\langle\psi_n^{(N)}|\chi_1\rangle\
\delta(E_n^{(N)}-E)\,. \label{dr}
\end{equation}
While $\Phi_N(\sigma)\rightarrow\Phi(\sigma)$ as $N$ tends to
infinity,
the amplitudes entering Eq. (\ref{dr})
are rather chaotic in the continuous spectrum region $E>E_{th}$,
and $r_N(E)$ thus does not tend to the true $r(E)$
in the normal sense.
However, if approximate solutions to the integral equation
are sought for
in the class of smooth functions then
the sequence of the approximate solutions $r_N^{smooth}(E)$ will
tend to the true $r(E)$
as $N$ goes to infinity.
(See also the next section.)

In other words, if one averages both $r(E)$ and $r_N(E)$ over $E$
using
a smooth kernel $K(\sigma,E)$ then the corresponding
"averaged $E$--overlaps"
$\Phi$ and $\Phi_N$  should
be close to each other
provided that $N$ is high enough. If $K$ effectively vanishes
for $E$ values
beyond some range
$\Delta E$ (depending on $\sigma$), the case most interesting
for us,
one can expect that the two averaged $E$--overlaps will be close
to
each other when
a sufficient number of levels $E_n^{(N)}$ fall into
$\Delta E$. Then one can use the averaged
discretized  $E$--overlap $\Phi_N(\sigma)$ to reconstruct the true
$E$--overlap $r(E)$
via solving Eq. (\ref{tra}) in an appropriate way.

The procedure is applicable with any smooth $K(\sigma,E)$, say
$K=\exp[-(\sigma-E)^2/\sigma_0^2]$. On the other hand, it is
restricted
to the case when
both $\chi_1$ and $\chi_2$ are localized. Besides, the complete
diagonalization of a Hamiltonian is not manageable for larger
systems.
Let us consider the
evaluation of the transform
with the help of Eq. (\ref{phi}) without the above
restriction and without having recourse to the
diagonalization of a
Hamiltonian. Unlike the considerations above, this evaluation
will be done for the kernels of
a special form. We set first
\begin{equation}
K(\sigma,E)=(E+\sigma)^{-1}\,.  \label{fo}
\end{equation}
According to Eq. (\ref{phi}), this leads to the transform
\begin{equation}
\Phi(\sigma)=\langle\chi_2|(\hat{H}+\sigma)^{-1}\chi_1\rangle\,.
\label{sti}
\end{equation}
Eq. (\ref{sti}) can be written  as
\begin{equation}
\Phi(\sigma)=\langle\chi_2|\tilde\Psi(\sigma)\rangle\,, \label{spc}
\end{equation}
where $\tilde\Psi$ is
the solution to the Schr\"odinger--like equation with a {\em source},
\begin{equation}
(\hat{H}+\sigma)\tilde{\Psi}=\chi_1\,.
\label{equ}
\end{equation}
 The $\sigma$
values we use are such that $-\sigma$ lie apart from the spectrum of
the Hamiltonian
so that $K$ is bounded. Due to this reason  $\tilde\Psi$ is
localized along with $\chi_1$
as it is explained above in connection
with Eq.  (\ref{appr}).  Due to
the same reason there exists only one localized solution to Eq. (\ref{equ}):
the difference of any two localized solutions would satisfy the homogeneous
equation that has no solutions yet.

One can also use
\begin{equation}
K(\sigma,E)=[(E+\sigma^*)(E+\sigma)]^{-1}\,. \label{2}
\end{equation}
We consider here the case of the response function (\ref{r}) and replace
$E$ with $\epsilon=E-E_0$ in Eq. (\ref{2}).
According to Eq. (\ref{phi}) the transform is
\begin{equation}
\Phi(\sigma)=(1/2)\left[
\langle\hat{O}_2\Psi_0|(\hat{H}-E_0+\sigma^*)^{-1}
(\hat{H}-E_0+\sigma)^{-1}\hat{O}_1|\Psi_0\rangle+c.c.\right]\,.
 \label{sord}
\end{equation}
The quantity (\ref{sord}) can be calculated as
\begin{equation}
\Phi(\sigma)=(1/2)\left[
\langle\tilde{\Psi}_2(\sigma)|\tilde{\Psi}_1(\sigma)\rangle+
\langle\tilde{\Psi}_1(\sigma)|\tilde{\Psi}_2(\sigma)\rangle\right]\,,
 \label{so}
\end{equation}
where $\tilde\Psi_i$ are solutions to
\begin{equation}
(\hat{H}-E_0+\sigma)\tilde{\Psi}_1=\hat{O}_1\Psi_0\,\,\,\,\,\,
(\hat{H}-E_0+\sigma)\tilde{\Psi}_2=\hat{O}_2\Psi_0\,.
\label{equ1}
\end{equation}
The first of Eqs. (\ref{equ1}) is, of course, Eq. (\ref{equ})
in the different notation.
From Eqs. (\ref{so}) at $\hat{O}_1=\hat{O}_2$
it could be inferred once more
that $\tilde{\Psi}_i$ are localized.
The general method we discuss has been given in Ref. \cite{efr85}
with use of the kernels
(\ref{fo}) and (\ref{2}) for real $\sigma$ values. The corresponding integral
transforms are called the Stieltjes and generalized Stieltjes
transform \s{wid}. We also note that if
Eqs. (\ref{equ1}) are solved
with the help of an expansion of $\tilde{\Psi}$ over a set of
known functions then e.g. the quantity (\ref{sord}) thus obtained coincides
with that
in Eq. (\ref{phin}) with the kernel (\ref{2}) provided that the
diagonalization of a
Hamiltonian is done in the subspace of retained basis functions.

So far we chose the kernel in a way that the 
evaluation of the transform, Eq. (\ref{phi}), is possible. 
Now let us discuss the choice of the kernel
from the point
of view of the stable reconstructing the response from the
integral equation
(\ref{tra}). At the same accuracy in $\Phi$, the inversion of the
transform
(\ref{tra}) will be the most stable if the kernel is chosen as
"narrow" as possible.
Indeed, at
integrating in Eq. (\ref{tra}), variations of $r$ occurring at
intervals $\Delta E$
that are smaller than the range of the kernel may
smear out and
even tend to cancel at some $\sigma$. It then may be hard to reconstruct
such changes
in presence of numerical inaccuracies in $\Phi$. The
kernels (\ref{fo})
and (\ref{2}) with real $\sigma$
do not possess a "finite" range and thus
are unfavorable from this point of view.
 Use of complex $\sigma =-\sigma_R+i\sigma_I$
in Eq. (\ref{2})
has been suggested in the
response function case \s{ELO94}
to increase the stability. The kernel then takes the "Lorentz" form, and
\be
\sum_n \frac{R_n}{(\sigma_R-\epsilon_n)^2+\sigma_I^2}+
\int_{\epsilon_{th}}^\infty\frac{R(\epsilon)}{(\sigma_R-
\epsilon)^2+\sigma_I^2}d\epsilon=\Phi(\sigma)\,.
\la{l2}
\ee
Here the
contribution
from discrete levels is written down explicitly.
For a stable reconstruction of the continuous part of the
response it is advantageous to use
mostly $\sigma_R\ge \epsilon_{th}$,
and $\sigma_I$ should be chosen "sufficiently small".
In some sense
$\sigma_I$ plays the role
of the energy resolution: any two responses that noticeably differ
from each other
at $\Delta \epsilon$
intervals not much smaller than $\sigma_I$ lead to noticeably
different transforms $\Phi$ and thus
can be discriminated even in presence of numerical uncertainties in
$\Phi$. But
in order to distinguish between responses differing  from each
other in regions small
compared to $\sigma_I$ a higher numerical accuracy in $\Phi$
than that required in $R$ might be necessary.

The procedure similar to that used above to calculate the
Lorentz transform in
the response function case can also be used for other $E$--overlaps
provided that in Eq. (\ref{gr}) both $\chi_1$ and $\chi_2$ are
localized. The following procedure can be suggested in the general case.
It is
applicable for breakup reactions with three or more fragments
in the final state when only
$\chi_1$ is localized.  Ons writes down the kernel (\ref{2}) as
\be
K(\sigma,E)=(2i\sigma_I)^{-1}[(E+\sigma^*)^{-1}-(E+\sigma)^{-1}]\,,
\la{K}
\ee
and, similar to Eq. (\ref{spc}), one has
\[
\Phi(\sigma)=(2i\sigma_I)^{-1}[\langle\chi_2|\tilde{\Psi}'\rangle
-\langle\chi_2|\tilde{\Psi}\rangle]\,,
\]
where $\tilde{\Psi}'$ and $\tilde{\Psi}$ are,
respectively, the localized solutions to the
equations
\be
(\hat{H}+\sigma^*)\tilde{\Psi}'=\chi_1\;\;\;
(\hat{H}+\sigma)\tilde{\Psi}=\chi_1\,. \la{ale}
\ee
In the response function case one may replace $E$ by
$\epsilon=E-E_0$, and $\hat{H}$ by $\hat{H}-E_0$
in Eqs. (\ref{K}) and (\ref{ale}), respectively.\footnote{It is easy to
conclude from Eqs. (\ref{K}), (\ref{ale}) that {\em in the response
function case} the "squared Lorentz" transform 
$K^2(\sigma,E)\equiv[(E-E_0+\sigma^*)(E-E_0+\sigma)]^{-2}$ is also calculable
in a simple form. The kernel $K^2$ is more sharp peaked than $K$ which 
facilitates the inversion.} 

The kernel (\ref{fo}) with complex $\sigma=-\sigma_R+i\sigma_I$:
\be K(\sigma,E)=[(E-\sigma_R)-i\sigma_I]/[(E-\sigma_R)^2+\sigma_I^2]
\la{cs}\ee
is also presumably good from the point of view of the
inversion stability.

We note that 
no information on exit channels of a reaction
is required in order to solve the above listed dynamic equations.
In the response function case, the long--range
Coulomb interaction manifests itself only in Eqs. (\ref{equ1}) and
does not cause any problems.

To calculate the response function, Eq. (\ref{r}), it is convenient
to use the multipole expansion of the transform. This allows taking into account contributions
from various projections of angular momenta in a general form. We  perform
the calculation in the reference system where the initial state
$\Psi_0$ is at rest
and carries a given total spin $J_0$ and its projection $M_0$.
It will be
denoted $\Psi_{J_0M_0}$ here. We set
$\hat{O}_1=\hat{O}_2=\hat{O}$ in Eq. (\ref{r}) and consider
the response $\bar{R}$ averaged over $M_0$ along with the
corresponding
transform $\bar{\Phi}$:
\be
\bar{R}=(2J_0+1)^{-1}\sum_{M_0=-J_0}^{J_0}R(M_0)\qquad
\bar{\Phi}=(2J_0+1)^{-1}\sum_{M_0-J_0}^{J_0}\Phi(M_0).            \la{bar}
\ee
We expand the operator $\hat{O}$ in a sum of irreducible tensor operators
$\hat{O}_{jm}$:
\be
\hat{O}=\sum_{jm}a_{jm}\hat{O}_{jm}, \la{ojm}
\ee
where $a_{jm}$ are the expansion coefficients. We define the partial
right--hand sides
\[
q_{JM}^j=\sum_{m+M_0=M}C^{JM}_{jmJ_0M_0}\hat{O}_{jm}\Psi_{J_0M_0},
\]
the partial solutions $\tilde{\Psi}_{JM}^j$:
\be
(\hat{H}-E_0+\sigma)\tilde{\Psi}_{JM}^j=q_{JM}^j, \la{pe}
\ee
and the partial transforms:
$\Phi_{jJ}=\langle q_{JM}^j|\tilde{\Psi}_{JM}^j\rangle$, or
$\Phi_{jJ}=\langle \tilde{\Psi}_{JM}^j|\tilde{\Psi}_{JM}^j\rangle$
for the case of Eq. (\ref{fo}) and Eq. (\ref{2}), respectively. The latter
quantities do not depend on $M$.  (These quantities are the
corresponding transforms of the components $R_{jJ}$ of response
functions which correspond to transitions to the continuum states with
given ${JM}$ induced by the $\hat{O}_{jm}$ multipole operators.)

One has
\[
\hat{O}\Psi_{J_0M_0}=\sum_{jmJM}a_{jm}C^{JM}_{jmJ_0M_0}q_{JM}^j\,\,\,\,\,\,\,
\tilde{\Psi}=\sum_{jmJM}a_{jm}C^{JM}_{jmJ_0M_0}\tilde{\Psi}_{JM}^j.
\]
Putting these expressions into Eq. (\ref{spc}) or (\ref{so}),
substituting this into the second of Eqs. (\ref{bar}), and then
performing the summation of the products of Clebsh--Gordan
coefficients one finally gets
\be
\bar{\Phi}=(2J_0+1)^{-1}\sum_{jJ}(2J+1)\Phi_{jJ}\sum_{m}(2j+1)^{-1}|a_{jm}|^2.
\la{me}
\ee
We also note that in Eq. (\ref{pe})
\[
\langle\varphi_{JM}|q_{JM}^j\rangle
=(2J+1)^{-1/2}(\varphi_J||\hat{O}_J||\Psi_{J_0})
\]
where $\varphi_{JM}$ is a (basis) state with given $J,M$ values,
and the right--hand side includes a reduced matrix element (defined as
in \s{var}).

If $\hat{O}({\bf q})$ is the charge operator then one has in Eq. (\ref{ojm})
$a_{jm}=Y_{jm}^*(\hat{{\bf q}})$,
and the sum
over $m$ in Eq.  (\ref{me}) equals $(4\pi)^{-1}$. Let us consider the
transversal current operator ${\bf J}_t({\bf q})$.  Eq.  (\ref{ojm}) should be
modified to
\[{\bf J}_t({\bf q})=\sum_{jm\lambda}{\bf Y}_{jm}^{\lambda
*}(\hat{{\bf q}})\hat{O}_{jm}^\lambda \]
where ${\bf Y}_{jm}^{\lambda }$ are vector
spherical harmonics \s{var}, and the $\lambda$ superscript signifies "electric"
or "magnetic". The quantities defined above acquire the additional
superscript $\lambda$, and Eq. (\ref{me}) takes the form \[
\bar{\Phi}=[4\pi(2J_0+1)]^{-1}\sum_{jJ\lambda}(2J+1)\Phi_{jJ}^\lambda. \]
Here the orthogonality property of vector spherical harmonics was used.

It was implied in the formulas of this section that the
transition operators
$\hat{O}_{1,2}$ in Eq. (\ref{r}) do not depend on the energy transfer $\epsilon$.
If a dependence on $\epsilon$ occurs it leads to a dependence of $\hat{O}_{1,2}$
on $\hat{H}$
after performing the closure. Then the above methods for calculating the
transform should be modified. In most of the cases the $\epsilon$ dependence
emerges only as a dependence of nucleon form factors on $\epsilon$. The
nucleon form factors can be factored out from a calculation, and the
formulas of the above type then become applicable. Often it is sufficient
to divide out only one general factor. For single--particle nuclear currents
this takes place if one neglects the $\epsilon$
dependence of the form factor ratio $G_E^n/G_E^p$ or uses the similarity of
the form factors $G_M^n$ and $G_M^p$. In case of zero isospin nuclei there is
no interference terms between
isovector and isoscalar components of the transition operator,
and the same
can be achieved without approximations by dividing out the
isovector and isoscalar nucleon
form factors.

\section{Inversion of the transform}

One could compare the calculated transform
of a response function with the transform of experimental
data on the response function \s{efr80,cs94,dobr}, and this does not
require inversion of the transform. However at such a comparison physics
might be obscured by the fact that 
parts of different nature of a response  are mixed up in the
transform.  In addition, with a given approximate Hamiltonian, often
the comparison of a low--energy part of a response with
experiment is only sensible. At the transform level, the comparison
is then obscured by high--energy contributions.
(These difficulties are less pronounced in case of sharply
peaked kernels. But for such kernels inversion is also an easier
task.) Therefore performing inversion is preferable, and
below we consider procedures for solving Eq. (\ref{tra}).

First we disregard discrete contributions to the solution and thus
consider the integral equation
\be
\int_{E_{th}}^\infty K(\sigma,E)r(E)dE=\Phi(\sigma). \la{ie}
\ee
Afterwards we shall remove this restriction. In the following we
assume that Eq. (\ref{ie}) has only one solution. This is obviously
true for the kernel (\ref{fo}). Let us
demonstrate that this is also the case for the Lorentz kernel. The difference
$\Delta r(E)$ of two possible solutions would satisfy the homogenous equation
of the form
\[ \int_a^\infty dx[(x-y)^2+\beta^2]^{-1}\Delta r(x)=0. \]
This
leads to
\be
\int_a^\infty dx\int_a^\infty dy [(x-y)^2+\beta^2]^{-1}\Delta
r(x)\Delta r(y)=0.\la{dint}
\ee
Using the representation
\[
(x^2+\beta^2)^{-1}= (2\beta)^{-1}\int_{-\infty}^\infty
dk\exp(-\beta |k|+ikx),\]
$\beta>0$, one can rewrite the left--hand side of
Eq. (\ref{dint}) as
\be
(2\beta)^{-1}\int_{-\infty}^\infty
dk\exp(-\beta|k|)|\varphi(k)|^2\la{lhs}
\ee
with
\[
\varphi(k)=\int_{-\infty}^\infty dx\exp(ikx)\theta(x-a)\Delta r(x).
\]
The quantity (\ref{lhs}) can equal to zero only if $\varphi(k)=0$ that
leads to $\Delta r=0$.

In the calculations performed so far
\s{efr85,ELO93,ELO94,MKOL95,ELO97A,ELO97B,ELO97C,ELO99,ELOT99,ELO98} the
inversion procedure was the following. The solution was sought for in the
form
\be
r(E)=\sum_{n=1}^Nc_n\varphi_n(E,\alpha) \la{exp}
\ee
where $\chi_n$ are
known functions forming a complete set.  They include non--linear parameters
$\alpha$.  If one substitutes the expansion (\ref{exp}) into the left--hand side
of Eq.  (\ref{tra}) one obtains
$\sum_{n=1}^Nc_n\bar{\varphi}_n(\sigma,\alpha)$ where
$\bar{\varphi}_n$ are transforms of the basis functions. The expansion
coefficients $c_n$ and the parameters $\alpha$ are obtained from the
best fit requirement
\be
\sum_{k=1}^K\left|\Phi(\sigma_k)-
\sum_{n=1}^Nc_n\bar{\varphi}_n(\sigma_k,\alpha)\right|^2={\rm min}.
\la{cr}
\ee
At fixed $\alpha$ this leads to a system of linear equations for
the expansion coefficients. (At large $N$ this system may become
ill--defined. Instead of solving it, the SVD algorithm
 for a direct minimization in (\ref{cr}), see e.g.
\s{nr}, may be applied.)

In the
Lorentz case,
see \s{ELO94,MKOL95,ELO97A,ELO97B,ELO97C,ELO99,ELOT99,ELO98}
the values $\sigma_k=-\sigma_R^k+i\sigma_I^k$ in (\ref{cr}) were
chosen such
that the $\sigma_R^k$ points covered the interval
$\epsilon_{th}\le\epsilon\le\epsilon_{max}$ where $\epsilon_{max}$
is such
that $R(\epsilon>\epsilon_{max})$ is already very small.
The $\sigma_R^k$
points covered also some interval with
$\epsilon_{min}\le\epsilon\le\epsilon_{th}$
aiming a better description
of the low--energy
part of the response.  The $\sigma_I$ values were chosen taking
into
account the following. When $\sigma_I$ is tending to zero
$K(\sigma,\epsilon)\rightarrow (\pi/|\sigma_I|)
\delta(\sigma_R-\epsilon)$, and
$\Phi(\sigma)\rightarrow (\pi/|\sigma_I|)R(\sigma_R)$. Hence for
sufficiently small $\sigma_I$
the required relative accuracy in $R$ is the same as in $\Phi$,
and no additional uncertainty in $R$ arises due to the inversion
procedure. On the other hand, the
smaller $\sigma_I$ is the harder is to achieve this accuracy
at solving
the dynamic equation (\ref{equ1}) with the bound--state type methods. Indeed,
at $\sigma_I\rightarrow 0$ the scattering regime is recovered.
For few--nucleon
systems use of $\sigma_I$ values which are comparable to the widths
of response functions and fall into
the range between several MeV and 20 MeV was found to be the most
convenient.\footnote{For e.g.  three--cluster $^{11}$Li response convenient
$\sigma_I$ values proved also be comparable to the width of the peak.}
(Of
course, the final results should not depend on $\sigma_I$ that can serve as a
possible check.) Values of $\sigma$ such that
$\sigma=-\sigma_R+i\sigma_I(\sigma_R)$ with $\sigma_I$ increasing
as $\sigma_R$ increases were found to yield good results, in particular.

Solutions to general equations of the form (\ref{ie}) are known to be unstable
with respect to high frequency oscillations. Indeed, if $r(E)$ is the
solution with an exact $\Phi$ then, for example, $r_a(E)=r(E)+\lambda\sin
Et$ is the solution with the right--hand side
\[ \Phi_a(\sigma)=\Phi(\sigma)+
\lambda\int_{E_{th}}^\infty dEK(\sigma,E)\sin Et.\]
At any $\lambda$ the variation $\Phi_a-\Phi$ of the right--hand side becomes
indefinitely small at sufficiently large $t$. However, the corresponding
variation $r_a-r=\lambda\sin Et$ of the solution  may take values as large as
$\lambda$.

Because of the above instability one should not seek for the exact
solution
of the equation with a given approximate right--hand side. Instead, as it is
well--known,
a regularization procedure should be applied at finding the
solution. The regularization suppresses a quickly oscillating component in
the solution with an
approximate right--hand side but do not influence much
a
slowly varying component in the solution.
If $r_a$ and $r$ are solutions with
right--hand sides $\Phi_a$ and $\Phi$, respectively, and the difference
$\Phi_a-\Phi$ is small then only a quickly oscillating component
in the
difference $r_a-r$ may have large amplitudes.  Due to this the
regularization guarantees the
closeness of a regularized "solution" to the
exact one.

When one looks for a solution in the form (\ref{exp}) the number
$N$ of
retained functions plays the role of a regularization parameter.
If
a right--hand side $\Phi_a$ is accurate enough then, normally, the accuracy
in the solution $r(E)$ first increases when $N$ increases, and the results
look as being "convergent" with respect to $N$. But at a further increase
in $N$ the "solution" obtained inevitably acquires unphysical oscillations.
The higher accuracy in $\Phi$ is, the higher $N$ values are at which the
oscillating regime starts, and the higher accuracy in $r$ achieved at
lower $N$ values is.

Stability of inversion results with respect to $N$ served both for choosing
$N$ values and as the main quality criterion of an approximate solution. To
check the results, responses obtained with different sets of basis functions
were compared to each other as well. (Such a comparison implies that
the stability ranges $\Delta N$ exist
for each of the sets). Stability with respect to an accuracy in the input $\Phi$,
was, of course, also checked. The quality criterion related to sum rules
was applied in addition.
The following sum rules should
approximately be satisfied by the solution $r(E)$:
\be
\int_{E_{th}}^\infty r(E)dE=\langle\phi_2|\phi_1\rangle,
\qquad
\int_{E_{th}}^\infty Er(E)dE=\langle\phi_2|\hat{H}|\phi_1\rangle\,.
\la{sr2}
\ee

Incorporation of the known threshold behavior of $r(E)$
into basis functions entering Eq. (\ref{exp}) increased the accuracy of
inversion. Other features of the solution such as
positions, or widths, of narrow resonances (if known)
should similarly be taken
into account for the same purpose.

In some cases the accuracy can also be enhanced in the following way
\s{ELO94}. Let us consider, for example, the longitudinal response
function (\ref{r}) for the $(e,e')$ reaction. It may behave in a rather
peculiar way at small $\epsilon$. The origin is that the
lowest $l$ multipole contributions $R_l$ to $R=\sum_l R_l$ have
maxima in the threshold region while  other multipoles exhibit maxima in
the quasielastic peak region. It is helpful to apply the whole procedure
separately to those lower $R_l$ and to the sum of all other $R_l$.
The reason for an increase in accuracy is that
with use of simple functions $\varphi_n$
in Eq. (\ref{exp}) one can better describe the behavior of those pieces of the
response function at sufficiently low $N$ values than that of the total
response function.

The accuracy of the inversion results has been assessed in
Ref. \s{ELO94}) using the two--nucleon d$(e,e')$ longitudinal
response
as an example. An approximate Lorentz
transform with a
several per cent oscillating error was taken as an input.
In Fig. 1 the response obtained via inversion of the transform
at $\sigma_I=5$ MeV is shown
(dashed curve) along with the exact response (solid curve)
calculated in the conventional way. The quality of the inversion is
excellent.

Regularization procedures different from that discussed above
to solve the general integral equation of
the first kind  were applied in the
literature, see e.g. \s{nr,phil,TA}. In our case the following one, see
\s{phil,TA}, seems to be promising. Let us use the quantity
\[
\rho(f_1,f_2)=\left\{\int_{\sigma_{min}}^{\sigma_{max}}
\left|f_1(\sigma)-f_2(\sigma)\right|^2d\sigma\right\}^{1/2}.
\]
as a measure of proximity between $f_1$ and $f_2$. Let $\Phi_a(\sigma)$
be an approximate right--hand side used as the inversion input. We suppose
that
\be
\rho(\Phi_a,\Phi)\simeq\delta. \la{del}
\ee
The procedure requires estimating $\delta$. For this purpose one can
proceed as follows in our case. Let $\nu$ be some parameter (e.g. the
number of basis functions at solving Eqs. (\ref{equ}), (\ref{equ1}))
that determines an accuracy in
$\Phi_a$ at a given $\sigma$ value:
\[ \Phi_a(\sigma)=f(\nu),\,\,\,\,\,\,\,\,\Phi(\sigma)=f(\infty). \]
One can estimate the difference $f(\nu)-f(\infty)$ and hence the quantity
(\ref{del}) performing the calculation at various $\nu$ values, then 
approximating
the results with some analytic formula, such as e.g.
$f(\nu)=f(\infty)+C\nu^{-\gamma}$, and fitting the parameters
of the formula that include
$f(\infty)$. It is supposed that inversion results are not sensitive
to a precision in $\delta$ which normally is the case.

There is no sense to solve Eq. (\ref{ie}) with an accuracy higher than
$\delta$. Therefore one seeks for the solution in the class of functions $r(E)$
satisfying the condition
\be
\rho(Kr,\Phi_a)\le\delta \la{cond1}
\ee
where $Kr$ is the short notation for the left--hand side of Eq. (\ref{ie}).
We consider $\Phi_a$ such that
$\int_{\sigma_{min}}^{\sigma_{max}}|\Phi_a(\sigma)|^2d\sigma>\delta^2$, so
that $r=0$ does not belong to the class (\ref{cond1}) of functions.
One wants to
find an approximation to the true $r(E)$ that is stable with respect to small
variations in $\Phi$. To this aim, the functional
\be
\Omega[r]=\int_{E_{th}}^\infty\left[q(E)\left|r(E)\right|^2+p(E)\left
|dr/dE\right|^2\right]dE \la{om}
\ee
is used, where $q$ and $p$ are non--negative functions. The function
$r(E)$ from the class
(\ref{cond1}) at which the functional $\Omega[r]$ takes its minimal
value
is chosen as the approximate solution.
(The minimal value does exist.) If one adds a quickly oscillating
component to the true $r(E)$ then the contributions to $\Omega[r]$ from both terms
in the right--hand side increase. Therefore the function $r(E)$
giving a
minimum to $\Omega[r]$ belongs to the "smoothest ones" among those
satisfying Eq. (\ref{cond1}). (In Eq. (\ref{om})
the simplest choices $q=1,p=0$, or $q=0,p=1$ are,
in general, reasonable.) It can be shown \s{TA} that if one takes a sequence of
$\delta$ values tending to zero then the corresponding sequence of
approximate solutions found in the above way would tend to the exact $r(E)$.
Thus at sufficiently small $\delta$ closeness between the
approximate $r(E)$
constructed as described here and the exact $r(E)$ is guaranteed.

It is easy to show \s{TA} that $r(E)$ giving a minimum to
$\Omega[r]$ satisfies the condition (\ref{cond1}) with the {\em equality sign}
\footnote{This can be seen as follows. Suppose that
$r=r_0$ gives a minimum
to $\Omega[r]$, and $\rho(Kr_0,\Phi_a)<\delta$. Then there exist
surroundings ($\rho_E(r,r_0)<\epsilon$)
of $r_0$ such that $\rho(Kr,\Phi_a)<\delta$ for any $r$ belonging to the
surroundings. For the function $r=\gamma r_0$, $|\gamma|<1$, belonging
to these surroundings one has
$\Omega[r]=\gamma^2\Omega[r_0]<\Omega[r_0]$.
Therefore the conditions $\Omega[r_0]={\rm min}$ and $\rho(Kr_0,\Phi_a)<\delta$
are incompatible.}:
\be
\rho(Kr,\Phi_a)=\delta. \la{cond2}
\ee
Therefore one can seek for the minimum of $\Omega[r]$ at the additional
condition (\ref{cond2}). This is a classical variational problem which can be
solved using the Lagrange multiplier method. One forms the functional
\be
\rho^2(Kr,\Phi_a)+\mu\Omega[r], \la{solv}
\ee
finds its minima $r_\mu$ at given $\mu$ values and, finally,
fixes $\mu$ from the equation (\ref{cond2}) with $r=r_\mu$. One can show
\s{TA} that the corresponding solution $\mu=\mu(\delta)$ to this
equation necessarily exists.

In our case it is convenient to seek for minima of the functional
(\ref{solv})
using the expansion of Eq. (\ref{exp}) type. At given $\mu$ this
leads to a system of linear equations for the expansion coefficients.
If one puts $\mu$ equal to zero this system of equations
turns to that following from
Eq. (\ref{cr}). In contrast to the case of Eq. (\ref{cr}),
in the present case
one
can retain as many basis functions in the expansion as desired. As above to increase
the accuracy of the inversion one should incorporate known features of a
solution into basis functions.

Since  the $E$--overlap (\ref{gr}) may include discrete contributions
$\sim\delta(E-E_n)$, the sum  of the form $\sum_nK(\sigma,E_n)r(E_n)$
is present  in general in the left--hand side of Eq. (\ref{ie}).  A good
strategy is to calculate this sum beforehand which leads to Eq. (\ref{ie})
with a modified right--hand side.

It is convenient  to calculate
the required $r(E_n)$
from the equation of (\ref{equ1}) type
\[
(\hat{H}-E)\tilde{\psi}=\chi_1
\]
with real $E$ ranging in the vicinity of the level $E_n$. Then
at $E\rightarrow E_n$
\[
\langle\chi_2|\tilde{\psi}(E)\rangle\simeq(E_n-E)^{-1}r(E_n). \]

 Another possibility applicable in the framework of the
method (\ref{cr}) is to solve the corresponding equation of Eq. (\ref{l2})
type directly with inclusion of the basis functions
$\delta(E-E_n)$ into the expansion (\ref{exp}).

\section{Some other non--conventional approaches}

In Refs. \s{a,a1} the following method for calculating elastic scattering
and ionization processes in atomic physics was suggested and used.
One calculates the $T$--matrix (\ref{tm}) for  a set of complex
energies $E$ and then performs an extrapolation to real $E$ values. This
leads to the dynamic equation of Eq. (\ref{equ}) form at a sequence
of $\sigma_I$ values tending to zero. Such a method probably
requires more labour to attain accurate results
then the method of integral transforms.
In contrast to what was said
above concerning Eq. (\ref{exp}),
within such a framework one cannot use
the known information about a solution in order to
improve accuracy.
Probably this is the reason for unsatisfactory
results obtained in \s{a1}
in the threshold region. Besides, while one needs the extrapolation
$\sigma_I\rightarrow 0$, difficulties in
obtaining accurate solutions to  Eq.  (\ref{equ}) with
bound--state methods increase as $\sigma_I$ decreases.

In Ref. \s{ish94} the following method for calculating
response functions is suggested. One can represent the
$\delta$ function in Eq. (\ref{r}) as
$-(\pi)^{-1}{\rm Im}(E-E_\gamma+i\eta)^{-1}$, and, correspondingly,
the quantity (\ref{r}) as (cf. (\ref{sti}), (\ref{spc}))
\begin{eqnarray}
-(\pi)^{-1}{\rm Im}\langle\hat{O}_2\Psi_0|\bar{\Psi}_1\rangle\,,
\nonumber\\
\bar{\Psi}_1=(E-\hat{H}+i\eta)^{-1}\hat{O}_1\Psi_0. \la{gl}
\eer
Here $E=E_0+\epsilon$. Like the method of integral transforms,
this procedure allows avoiding summation over final states
at the calculation of a response function. However, in contrast to
the state (\ref{equ}), or (\ref{equ1}),
the state (\ref{gl}) is not localized, and at large
distances it includes
outgoing waves in all open channels. The continuum-state calculation
techniques are necessary to find this state which requires
more effort then obtaining the localized state (\ref{equ}). The method was
successfully applied for calculating three--nucleon responses
\s{ish94,glo}.

\section{Few--nucleon responses from the Lorentz transform}

These responses are smooth functions with single peaks and without
pronounced additional structures which facilitates calculations.
In Ref. \s{MKOL95} the longitudinal three--nucleon $(e,e')$
response functions were studied. Eq. (\ref{equ1}
(for $\hat{O}_1=\hat{O}_2)$ was cast into the form of inhomogenious
Faddeev--like equations, and the source term was split into the three
parts in a way preserving the permutational symmetries of Faddeeev
components. Similar to Ref. \s{ELO94}, inversion was performed separately
for the monopole contribution and for the sum of all other
contributions to the response functions. The responses thus obtained
compared well with those obtained from a conventional lengthy
calculation, however the second above mentioned piece of the
response function exhibited a small unphysical wiggle at low energy.
When the wiggle was removed "by hand" the comparison with the
conventional calculation became perfect. Probably one could have
come to the same results if the proper threshold behavior of
that piece of the response ($(\epsilon-\epsilon_{th})^{3/2}$ in the
$^3$H case) had been imposed.
It was demonstrated that the
final results are nearly independent of $\sigma_I$, as they should.

In Refs. \s{ELO97A,ELO97B,ELO97C,ELO99,ELOT99,ELO98}
the
three-- and four--nucleon responses were obtained
using an expansion technique in the dynamic equation
(\ref{equ1}) ($\hat{O}_1=\hat{O}_2$). The $^4$He$(e,e')$ longitudinal
response function
\s{ELO97A}, the total $^4$He \s{ELO97B}, and tri--nucleon
\s{ELO97C,ELO99,ELOT99}, photodisintegration cross sections,
and the $^4$He spectral function along with the
corresponding approximate
$^4$He$(e,e')$ response
\s{ELO98} were studied. Both central semi--realistic
\s{ELO97A,ELO97B,ELO97C,ELO98} and realistic \s{ELO99,ELOT99}
NN interactions supplemented with a 3N force \s{ELOT99} were
used. In \s{ELO97A,ELO97B,ELO97C,ELO99,ELO98} the correlated
hyperspherical expansion \s{fe72} was applied to solve the problem.
The basis functions were taken as products of the Jastrow factor,
 properly symmetrized hyperspherical harmonics
coupled with the corresponding spin--isospin functions,
and hyperradial
functions. In Ref.
\s{ELOT99} the basis function are constructed as above mentioned
products of spin--isospin--hyperspherical and hyperradial functions
to which a Jastrow type spin--isospin dependent operator is
applied. The prescription for selection of basis
hyperspherical harmonics \s{efr72},
used in older work with no Jastrow correlations, is applied
in conjunction with these correlations and is found to be
very effective both for bound states and at solving Eq. (\ref{equ1})
\s{ELOT99}.

It was found that the calculation of the transforms with an accuracy
at a per cent level provides stable response functions.
Achievment of such an accuracy in the transforms proved to be not
requiring more effort than the calculation of binding energies
with a comparable accuracy. As an example, in Fig.2 the dependence
of the Lorentz transform on the maximal value of the
hyperspherical
number $K$ retained in the calculation is shown in the case
of the four--nucleon photodisintegration
response function. The calculation was done with the
semi--realistic Malfliet--Tjon potential. The transforms at
$K_{max}=7$ and at $K_{max}=5$ are already very close to each other.
The sum rule checks (\ref{sr2}) were
satisfied  by the responses obtained
with a per cent accuracy. At the same
time, let us mention
that all possible checks of inversion results should be done,
while the criterion of intermediate stability of the results with
respect to the
number  $N$ of basis functions in the expansion (\ref{exp}) may
prove to be
insufficient by itself \s{ELO99}.
Convergent 3--nucleon photodisintegration
results with a realistic NN interaction supplemented with a 3N force
are presented in \s{ELOT99}.

Finding practical ways to perform the calculations for
larger systems, applying the approach to exclusive processes and
to responses with more complicated structures,
taking into account the long--range Coulomb interaction for
exclusive reactions
remains for
future.

The author wishes to thank W. Leidemann and G. Orlandini, his
collaborators in the development of the present approach.
The remarks on discretized responses 
were initiated by common calculations with I. Thompson.
The work was supported by the Russian Foundation for Basic
Research (grants no 96--15--96548 and 97--02--17003).

\begin{figure}The d$(e,e')$ longitudinal response function. Solid
curve: the conventional two--nucleon calculation. Dashed curve:
calculation from the inversion of the Lorentz integral transform
($\sigma_I=5$ MeV).
\end{figure}
\begin{figure}The Lorentz transform
$L(\sigma=-\sigma_R+i\sigma_I(\sigma_R))$
of the $^4$He photodisintegration
response function at various $K_{max}$ values. The calculations
were done for the Malfliet--Tjon NN potential. The $\sigma_I$ value
ranged from 5 MeV to 20 MeV as $\sigma_R$ increased.
\end{figure}
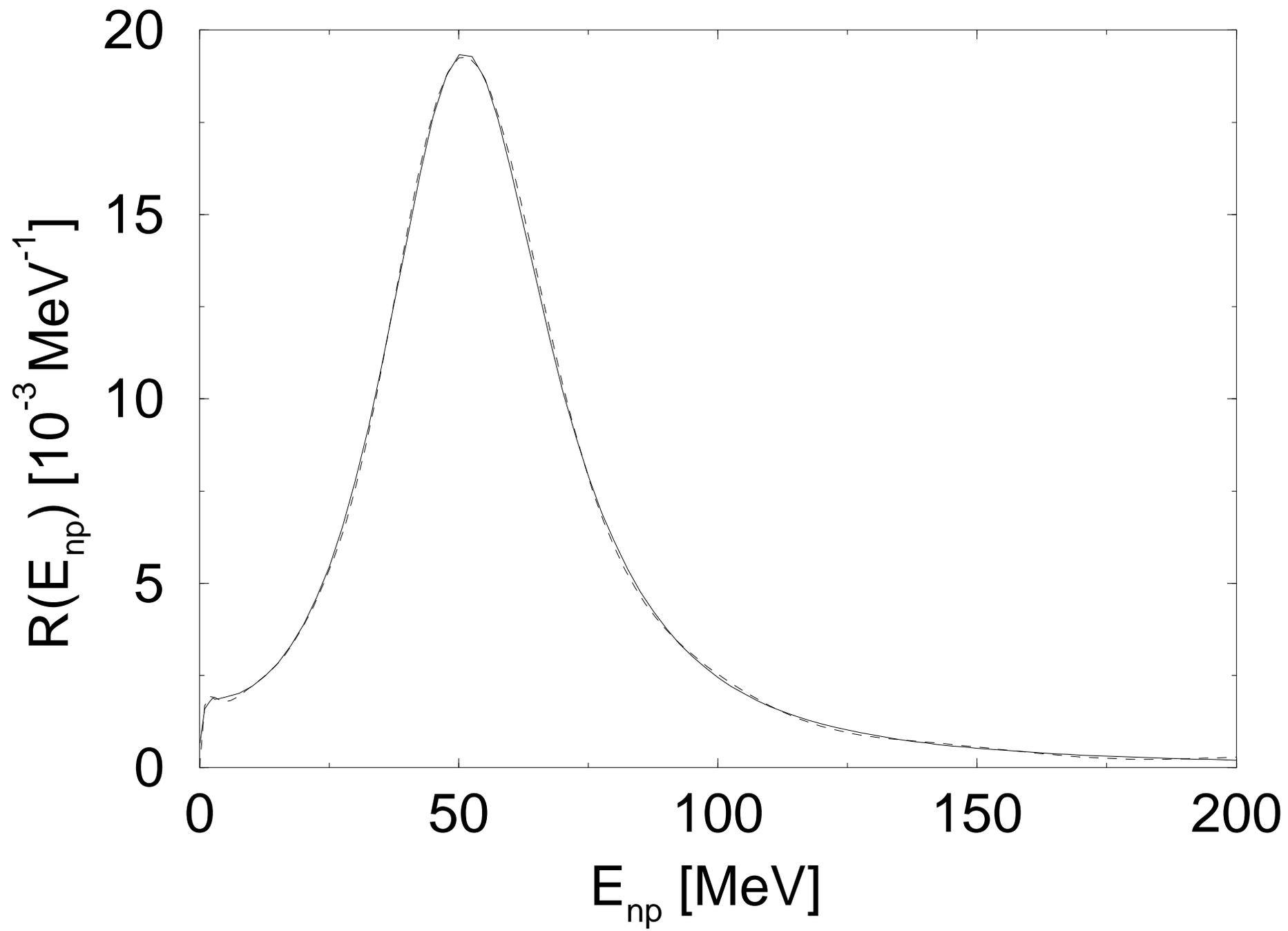
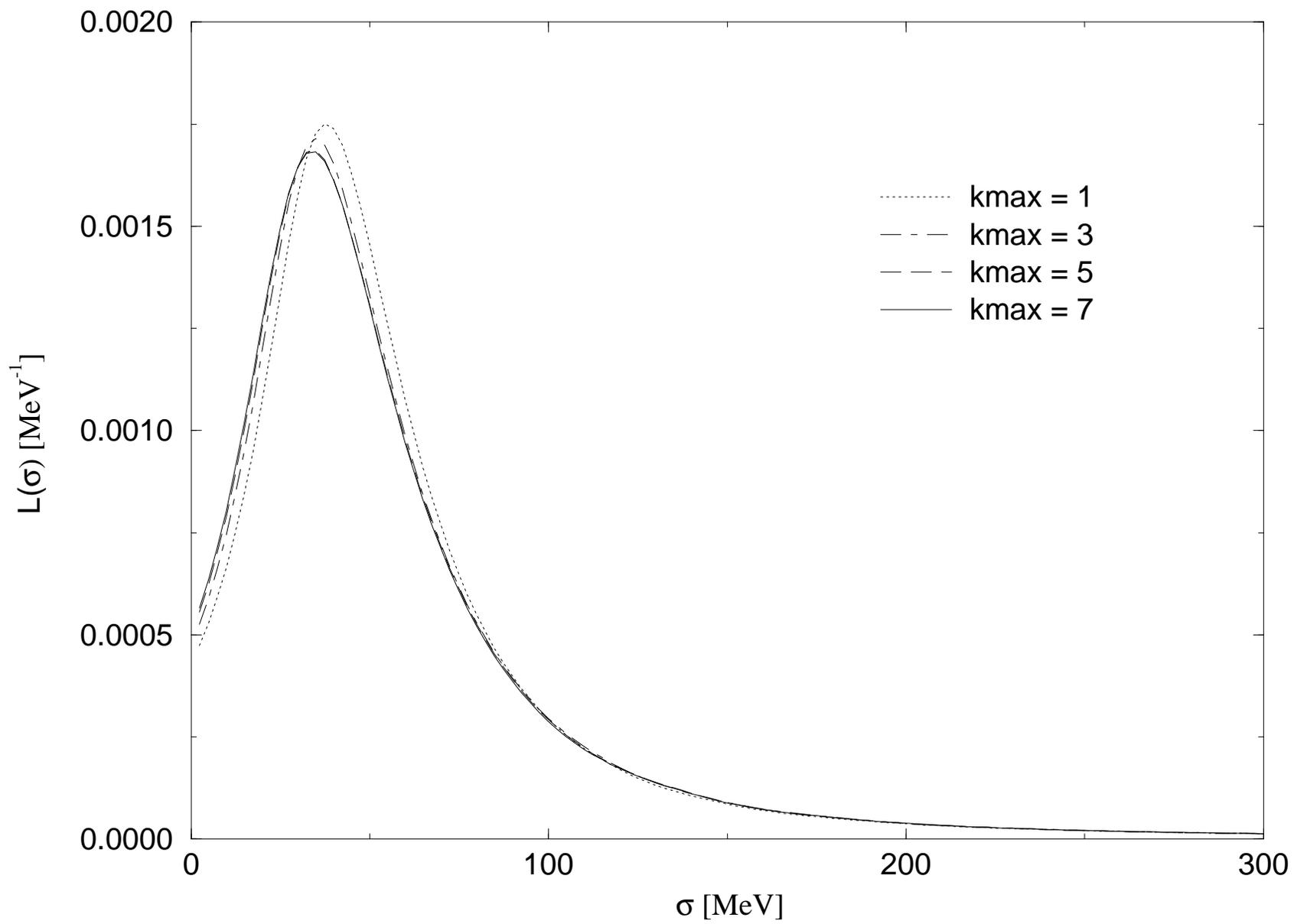
\end{document}